\documentclass[11pt]{article}

\usepackage{graphicx,amsmath,amsfonts,amssymb,dcolumn,amsthm,slashed,bbm,xspace,pgffor,tikz,mathbbol,latexsym,enumerate}
\usepackage{soul}
\usepackage[colorlinks,citecolor=red,urlcolor=cyan,linkcolor=blue]{hyperref}
\usepackage[utf8]{inputenc}
\usepackage[english]{babel}
\usepackage{lmodern}
\usepackage{microtype}
\usepackage{cite}

\numberwithin{equation}{section}

\begin{document}

\title{\textbf{Carroll limit of four-dimensional gravity theories in the first order formalism}}

\author{\textbf{Amanda Guerrieri¹}\thanks{amguerrieri@cbpf.br} \ and \textbf{Rodrigo F.~Sobreiro²}\thanks{rodrigo\_sobreiro@id.uff.br}\\\\
¹\textit{{\small CBPF - Centro Brasileiro de Pesquisas Físicas,}}\\
\textit{{\small R. Dr. Xavier Sigaud, 150, 22290-180, Rio de Janeiro, RJ, Brasil.}}\\
\\
²\textit{{\small UFF - Universidade Federal Fluminense, Instituto de F\'isica,}}\\
\textit{{\small Av. Litorânea, s/n, 24210-346, Niter\'oi, RJ, Brasil.}}}

\date{}
\maketitle

\begin{abstract}
We explore the ultra-relativistic limit of a class of four dimensional gravity theories, known as Lovelock-Cartan gravities, in the first order formalism. First, we review the well known limit of the Einstein-Hilbert action. A very useful scale symmetry involving the vierbeins and the boost connection is presented. Moreover, we explore the field equations in order to find formal solutions. Some remarkable results are obtained: Riemann and Weitzenb\"ock like manifolds are discussed; Birkhoff's theorem is verified   for the torsionless case; an explicit solution with non-trivial geometry is discussed; A quite general solution in the presence of matter is obtained. Latter, we consider the ultra-relativistic limit of the more general Lovelock-Cartan gravity. The previously scale symmetry is also discussed. The field equations are studied in vacuum and in the presence of matter. In comparison with the Einstein-Hilbert case, a few relevant results are found: Birkhoff's theorem is also verified   for the torsionless case; A quite general solution in the presence of matter is obtained. This solution generalizes the previous case; Riemann and Weitzenb\"ock like manifolds are derived in the same lines of the Einstein-Hilbert case.
\end{abstract}

\newpage

\tableofcontents

\newpage

\section{Introduction}

General Relativity (GR) \cite{Misner:1974qy,Wald:1984rg,DeSabbata:1986sv} governs the motion of particles near massive compact macroscopic objects by deforming the spacetime around it. Moreover, GR is also responsible to describe the causal structure of spacetime. When we are dealing with the weak field limit, particles upon this geometry are traveling at low velocities and we can describe their motion by Newton's theory of gravity, a non-relativistic weak field limit of GR. The covariant version of this regime is typically known as Newton-Cartan or Galilei gravity, see for instance \cite{Cartan:1923zea,Cartan:1924yea,Trautman:1963aaa,Havas:1964zza,Trautman:1965aaa,Kunzle:1972aaa,Dixon:1975fy,Banerjee:2016laq,Bergshoeff:2017dqq,Hansen:2020pqs,Guerrieri:2020vhp}. In contrast, it is also possible to take the ultra-relativistic (UR) limit of gravity and describe the motion of particles with extremely high energies traveling very close to the speed of light, at a strong field regime. Such regime is known as Carroll gravity, see \cite{Hartong:2015xda, Bergshoeff:2017btm,Ciambelli:2018xat, Bergshoeff:2014jla, Bergshoeff:2020xhv}. 

One efficient and systematic way to attain such limits in a gravity theory is to look to some inherent symmetries of gravity. Perhaps the most interesting symmetry to exploit is the local spacetime isometries because they are directly related to the equivalence principle. Moreover, the local isometries ensure a gauge theoretical character for gravity. In fact, gravity can be thought as a gauge theory of the the Poincar\'e group\footnote{More precisely, only the Lorentz sector is gauged, since the Poincar\'e group is not a semi-simple Lie group due to the orthogonal Abelian annex of translations. Nevertheless, the translational sector is still present since it is defines a fundamental representation of the Lorentz group.}, describing all local inertial frames \cite{Utiyama:1956sy,Kibble:1961ba,Sciama:1964wt,Mardones:1990qc,Zanelli:2005sa}. Such description of gravity is also known as first order formalism of gravity, since the field equations are of first order in the derivatives of the fundamental geometrical fields, \emph{i.e.}, the vierbein and the Lorentz connection. In this scenario, Galilei and Carroll gravities are obtained from suitable contractions of the Poincar\'e group known as Inönü-Wigner contractions \cite{Inonu:1953sp}. Galilei gravity is obtained considering the limit where the relative velocities between the local inertial frames are small in comparison with the light speed \cite{Cartan:1923zea,Cartan:1924yea,Trautman:1963aaa,Havas:1964zza,Trautman:1965aaa,Kunzle:1972aaa,Dixon:1975fy,Banerjee:2016laq,Bergshoeff:2017dqq,Hansen:2020pqs,Guerrieri:2020vhp}. The result is a gauge theory for the Galilei group. Under suitable assumptions and additional information \cite{Christensen:2013rfa,Banerjee:2014nja,Afshar:2015aku,Abedini:2019voz}, Galilei gravity is equivalent to Newton's theory of gravity. Hence, the Galilei limit assumes the full opening of the light cones. 

On the other hand, Carroll gravity arises by taking the relative velocities between frames to approach to the light speed while the light speed tends to zero \cite{Hartong:2015xda, Bergshoeff:2017btm,Ciambelli:2018xat, Bergshoeff:2014jla, Bergshoeff:2020xhv}. Therefore, the Carroll limit assumes the full closing of the light cones. The result is a gauge theory for the Carroll group \cite{Leblond, Bacry1968, Duval:2014lpa}. Interestingly, Carrollian spacetimes describe the geometry of null hypersurfaces in Lorentzian spacetimes defined with an extra dimension \cite{Ciambelli:2019lap}. It was  noticed in \cite{Barducci:2018wuj} that, in Carroll gravity, no interactions between spatially separated events occur. However, when isolated, these objects, initially viewed as immobile, show evolution in time (This effect is known as Carroll causality). For this reason, the Carroll limit is also known as the ultra local approximation of gravity. Recently, it was discovered that the UR gravity has properties associated with the strong force \cite{Fokas:2019zvp} and Carroll algebra plays an important role in flat space holography \cite{Ciambelli:2018wre, Duval:2014uva} and in Bondi-van der Burg-Metzner-Sachs symmetry \cite{Bondi:1962px,Sachs:1962zza,Duval:2014uva,Grumiller:2017sjh}.

In this work we proceed in two parts. In the first part, we study the Carroll limit of the Einstein-Hilbert (EH) action in the first order formalism and study the field equations at formal level. Thence, we consider the four dimensional Mardones-Zanelli (MZ) action \cite{Mardones:1990qc} describing Lovelock-Cartan (LC) gravity, which is just Lovelock's gravity improved with torsional terms. LC gravity is also a first order theory. In fact, in any spacetime dimension, MZ actions are  polynomially local in the fields and their derivatives, locally Lorentz invariant and explicitly metric independent. Thence, the Carrol limit of the MZ action and the corresponding field equations are explored at formal level.

In the case of the UR limit of EH action, we were able to find some novel results. First of all, we identify the emergence of a global scale symmetry similar to the accidental local scale symmetry of Galilei gravity \cite{Guerrieri:2020vhp}.   By imposing this symmetry to the matter content, a restricted form for the matter action coupled to Carroll gravity arises and facilitates the analysis. Hence, we are able to find a quite general solution for the curvatures and torsions in the presence of matter. Latter, we find the constraints on the matter content in order to the theory accept Riemannian-like geometries (Vanishing torsions and non-vanishing curvatures). The same analysis is performed for Weitzenböck-like manifolds (Vanishing curvatures and non-vanishing torsions). We also develop a non-trivial solution where space curvature and time torsion are the only non-trivial field strengths. In that particular example, we were able to compute the lapse and proper time as functions of the coordinate time. Finally, we confirm the validity of Birkhoff's theorem in the   torsionless Carroll limit of the EH action.

The case of the UR limit of MZ action is then considered in the second part of the paper. A generalized Carroll action, called here by Carroll-Cartan gravity, is obtained and the corresponding field equations are derived. The first property of the Carroll-Cartan gravity we find is that the scale symmetry is not present anymore due to the torsional terms. The symmetry, however, can be restored if we extend it to include extra transformations for the torsional coupling parameters. The matter action is assumed to be of the same form of the Carroll case. In   torsionless vacuum, Birkhoff's theorem is shown to be valid again. In the presence of matter, a quite general solution is obtained as a generalization of the general solution obtained for the Carroll case. Finally, we show that Riemannian-like and Weitzenböck-like spacetimes are also acceptable in Carroll-Cartan gravity.

  The models and results obtained here have many possible applications for future analysis. As mentioned, flat space holography \cite{Ciambelli:2018wre, Duval:2014uva} and Bondi-van der Burg-Metzner-Sachs symmetry \cite{Bondi:1962px,Sachs:1962zza,Duval:2014uva,Grumiller:2017sjh} are possibilities. Applications from gravity to solid state physics could follow, for instance, the lines of \cite{Isham:1975ur,Henneaux:1979vn}. Moreover, direct applications to actual gravitational systems within extremely high energy scales are also in our future perspective. In fact, the study of the Carroll-Cartan gravity is here justified because it can be implemented in such systems as effective theories where torsional terms account for extra possible effects. 

The paper is organized as follows: In Section \ref{LC} we construct the LC theory of gravity restricting ourselves to four dimensions. In Section \ref{IW} we obtain the Carroll group from a IW contraction of the Poincaré group and implement the contraction effects on the fields. In Section \ref{EH}, Carroll gravity is explored at formal level. Then, in Section \ref{GC}, Carroll-Cartan gravity is studied. Finally, our conclusions are displayed in Section \ref{FINAL}.

\section{Lovelock-Cartan gravity}\label{LC}

Our starting point is the MZ action \cite{Mardones:1990qc,Zanelli:2005sa}, describing a gravity theory over a four-dimensional Riemann-Cartan manifold (the spacetime) $M$,
\begin{eqnarray}
    S_{MZ}&=&\kappa\int\;\epsilon_{ABCD}\left(R^{AB}e^Ce^D+\frac{\Lambda}{2}e^Ae^Be^Ce^D\right)+\int\left(z_1R^{AB}e_Ae_B+z_2T^AT_A\right)+\nonumber\\
    &+&\int\left(z_3\epsilon_{ABCD}R^{AB}R^{CD}+z_4R^{AB}R_{AB}\right)+S_m\;.\label{mz1}
\end{eqnarray}
In the action \eqref{mz1}, $\kappa=1/8\pi G$, with $G$ being the Newton constant and $\Lambda$ is recognized as the cosmological constant. The constants $z_1$, $z_2$, $z_3$, and $z_4$ are free parameters with no correspondence with GR. The matter content coupled to gravity is denoted by $S_m$. The Lorentz indices (frame indices), denoted by Latin capital letters, run through $A,B,C\ldots\in\{$\underline{0},$\underline{1},$\underline{2},$\underline{3}\}$ (underlined numbers will denote frame indices). and can be raised and lowered with the help of the local Minkowski metric $\eta_{AB}=\eta^{AB}\equiv\mathrm{diag}(-,+,+,+)$. The totally antisymmetric object $\epsilon_{ABCD}$ stands for the Levi-Civita symbol in four dimensions. The fields $e^A$ and $e_A$ stands for the vierbein 1-form and its inverse. The 2-form fields $R^{AB}$ and $T^A$ are curvature and torsion, respectively given by,
\begin{eqnarray}
R^{AB}&=&d\omega^{AB}+\omega^A_{\phantom{A}C}\omega^{CB}\;,\nonumber\\
T^A&=&\nabla e^A\;\;=\;\;de^A+\omega^A_{\phantom{A}B}e^B\;,\label{2forms0}
\end{eqnarray}
with $\omega^{AB}=-\omega^{BA}$ being the Lorentz connection while $\nabla$ is the Lorentz covariant derivative. The Bianchi identities are easily derived,
\begin{eqnarray}
T^A&=&\nabla e^A\;,\nonumber\\
\nabla T^A&=&R^A_{\phantom{A}B} e^B\;,\nonumber\\
\nabla R^A_{\phantom{A}B}&=&0
\;.\label{hier0}
\end{eqnarray}

The vierbein and its inverse obey the following relations
\begin{eqnarray}
e^A_\mu e^\mu_B&=&\delta^A_B\;,\nonumber\\
e^A_\mu e_A^\nu&=&\delta_\mu^\nu\;,\label{inv0}
\end{eqnarray}
with lower case Greek indices (world indices) running through $\alpha,\beta,\gamma\ldots\in\{0,1,2,3\}$. The vierbein naturally induces a metric $g_{\mu\nu}$ (and its inverse $g^{\mu\nu}$) in $M$ through the relations:
\begin{eqnarray}
g_{\mu\nu}&=&e^A_\mu e^B_\nu\eta_{AB}\;,\nonumber\\
g^{\mu\nu}&=&e_A^\mu e_B^\nu\eta^{AB}\;,\nonumber\\
\eta^{AB}&=&e^A_\mu e^B_\nu g^{\mu\nu}\;,\nonumber\\
\eta_{AB}&=&e_A^\mu e_B^\nu g_{\mu\nu}\;.\label{metrics0}
\end{eqnarray} 

The action \eqref{mz1} is actually the most general gravity action in four dimensions which is polynomially local, explicitly metric independent, depending only on first order derivatives, and gauge invariant under infinitesimal $SO(1,3)$ local Lorentz gauge transformations of the form\footnote{The action \eqref{mz1} is also invariant under finite gauge transformations. Nevertheless, we will not use such transformations in the present study.}
\begin{eqnarray}
\delta\omega^{AB}&=&\nabla\alpha^{AB}\;,\nonumber\\
\delta e^A&=&\alpha^A_{\phantom{A}B}e^B\;.\label{gt0}
\end{eqnarray}
with $\alpha^{AB}=-\alpha^{BA}$ being an infinitesimal local parameter.

Finally, we identify each term in the action \eqref{mz1}: The first terms are, clearly, the EH action followed by the cosmological constant term. The terms in $z_1$ and $z_2$ are essentially equivalent, up to a surface term. Moreover, $z_1$ and $z_2$ have mass squared dimension.  The last two terms are of topological nature. In fact, the term in $z_3$ is the Gauss-Bonnet action and the term in $z_4$ is recognized as the Pontryagin term \cite{Mardones:1990qc,Zanelli:2005sa,Kobayashi,Nakahara:1990th}. Thence, these two last terms do not contribute to the field equations and the parameters $z_3$ and $z_4$ are dimensionless topological parameters. At last, it is interesting to notice that in the particular case when $z_2=-z_1$, the related terms become the Nieh-Yan topological term \cite{Mardones:1990qc,Zanelli:2005sa,Nieh:1981ww,Nieh:2007zz,Nieh:2018rlg}. It is worth mentioning that the MZ action generalizes the four-dimensional Lovelock gravity \cite{Lovelock:1971yv} by including torsional terms in the action.

The field equations can be easily derived for the fundamental fields $e^A$ and $\omega^{AB}$, providing,
\begin{eqnarray}
\kappa\epsilon_{ABCD}\left(R^{BC}e^D+\Lambda e^Be^Ce^D\right)+(z_1+z_2)R_{AB}e^B&=&-\frac{1}{2}\frac{\delta S_m}{\delta e^A}\;,\nonumber\\
\kappa\epsilon_{ABCD}T^Ce^D+\frac{(z_1+z_2)}{2}\left(T_Ae_B-e_AT_B\right)&=&-\frac{1}{2}\frac{\delta S_m}{\delta\omega^{AB}}\;.\label{feq0}
\end{eqnarray}
It can be verified by simple calculations that the typical asymptotic vacuum solution of these equations is a torsionless maximally symmetric spacetime \cite{Guerrieri:2020vhp}
\begin{eqnarray}
R^{AB}_0&=&-\Lambda e^Ae^B\;,\nonumber\\
T_0^A&=&0\;.\label{dS0}
\end{eqnarray}
Clearly, such solution leads to de Sitter or anti-de Sitter spacetimes, depending on the sign of the cosmological constant.

An important property of the LC gravity \eqref{mz1} is the validity of Birkhoff`s theorem \cite{Wald:1984rg}. To show that, one sets vanishing torsion and imposes a spherically symmetric form for the line element. Looking at the field equations \eqref{feq0} in vacuum, vanishing torsion automatically satisfies the second equation. Moreover, due to the second Bianchi identity in \eqref{hier0}, the term proportional to $(z_1+z_2)$ in the first equation also vanishes. Hence, at the level of the field equations, vanishing torsion implies on the reduction of LC gravity to the EH gravity with cosmological constant. It follows that Birkhoff's theorem is thus valid, providing a static Schwarzschild-de Sitter geometry.   In fact, as pointed out in \cite{Obukhov:2020hlp}, this solution is unique, even for a larger class of gravity theories. Nevertheless, if torsion does not vanish, a different spherically symmetric solution may arise. Such analysis for the present theory may be long and we leave it for a separated paper. For further generalizations of Birkhoff's theorem we also refer to \cite{Ramaswamy:1979zz,Neville:1979fk,Rauch:1981tva,Obukhov:1987tz,delaCruz-Dombriz:2018vzn} and references therein.

\section{Poincar\'e and Carroll algebras}\label{IW}

In this section, we discuss the UR limit of the Poincar\'e group in order to obtain the Carrol group. In fact, the gauge symmetry \eqref{gt0} of the MZ action \eqref{mz1} can be described by the Poincar\'e group $ISO(1,3)=SO(1,3)\times\mathbb{R}^{1,3}$, instead of the smaller Lorentz group $SO(1,3)$. Such description is allowed because the translational generators $\Pi_A$ of the sector $\mathbb{R}^{1,3}$ are also generators of the Lorentz group in the fundamental representation. The Lorentz sector itself has generators denoted by $L_{AB}$, with $L_{AB}=-L_{BA}$. The Poincar\'e algebra is then given by
\begin{eqnarray}
\left[L_{AB},L_{CD}\right]&=&\frac{1}{2}\left(\eta_{AD}L_{BC}-\eta_{AC}L_{BD}+\eta_{BC}L_{AD}-\eta_{BD}L_{AC}\right)\;,\nonumber\\
\left[L_{AB},\Pi_C\right]&=&\frac{1}{2}\left(\eta_{BC}\Pi_A-\eta_{AC}\Pi_B\right)\;,\nonumber\\
\left[\Pi_A,\Pi_B\right]&=&0\;.\label{poincalg1}
\end{eqnarray}

The first step towards the UR limit of the algebra \eqref{poincalg1} is to decompose the Poincar\'e group into space and time sectors, namely $ISO(1,3)=SO(3)\times L(3)\times\mathbb{R}_s^3\times\mathbb{R}_t$. Obviously, $\mathbb{R}_s^3$ stands for spatial translations and $\mathbb{R}_t$ for time translations. Thence,
\begin{eqnarray}
L_{AB}&\equiv&\left(L_{ab},L_{a\underline{0}}\right)\;\;=\;\;\left(L_{ab},L_a\right)\;,\nonumber\\
\Pi_A&\equiv&\left(\Pi_a,\Pi_{\underline{0}}\right)\;\;=\;\;\left(\Pi_a,\Pi\right)\;,\label{poincalg2}
\end{eqnarray}
where lowercase Latin indices run through $a,b,c\dots h\in\{\underline{1},\underline{2},\underline{3}\}$. Hence, the Poincar\'e algebra \eqref{poincalg1} decomposes as
\begin{eqnarray}
\left[L_{ab},L_{cd}\right]&=&\frac{1}{2}\left(\delta_{ad}L_{bc}-\delta_{ac}L_{bd}+\delta_{bc}L_{ad}-\delta_{bd}L_{ac}\right)\;,\nonumber\\
\left[L_{ab},L_c\right]&=&\frac{1}{2}\left(\delta_{bc}L_a-\delta_{ac}L_b\right)\;,\nonumber\\
   \left[L_a,L_b\right]&=&\frac{1}{2}L_{ab}\;,\nonumber\\
\left[L_{ab},\Pi_c\right]&=&\frac{1}{2}\left(\delta_{bc}\Pi_a-\delta_{ac}\Pi_b\right)\;,\nonumber\\
\left[L_a,\Pi_b\right]&=&-\frac{1}{2}\delta_{ab}\Pi\;,\nonumber\\
   \left[L_a,\Pi\right]&=&-\frac{1}{2}\Pi_a\;,\label{poincalg3}
\end{eqnarray}
and zero for all other commutators. Consequently, the algebra-valued forms follow the same decomposition, namely
\begin{eqnarray}
e^A\Pi_A&=&e^a\Pi_a+q\Pi\;,\nonumber\\
\omega^{AB}L_{AB}&=&\omega^{ab}L_{ab}+\theta^aL_a\;,\nonumber\\
T^A\Pi_A&=&T^a\Pi_a+\mathcal{Q}\Pi\;,\nonumber\\
R^{AB}L_{AB}&=&\Omega^{ab}L_{ab}+S^aL_a\;,\label{fdecomp1}
\end{eqnarray}
  with $q=e^{\underline{0}}$, $\theta^a=2\omega^{a\underline{0}}$, $\mathcal{Q}=T^{\underline{0}}$, and $S^a=2R^{a\underline{0}}$. Moreover,
\begin{eqnarray}
   T^a&=&De^a+\frac{1}{2}q\theta^a\;,\nonumber\\
\mathcal{Q}&=&dq-\frac{1}{2}\theta_ae^a\;,\nonumber\\
   \Omega^a_{\phantom{a}b}&=&R^a_{\phantom{a}b}+\frac{1}{4}\theta^a\theta_b\;,\nonumber\\
S^a&=&D\theta^a\;,\label{fdecomp2}
\end{eqnarray}
where the covariant derivative $D$ is assumed to be taken with respect to the $SO(3)$ sector by means of $D\cdot^a= d\cdot^a+\omega^a_{\phantom{a}b}\cdot^b$, and we have defined $R^{ab}=d\omega^a_{\phantom{a}b}+\omega^a_{\phantom{a}c}\omega^c_{\phantom{c}b}$. The nomenclature of all fields are now in order: $q$ - \emph{time vierbein}; $e$ - \emph{space vierbein}; $\theta$ - \emph{boost connection}; $\omega$ - \emph{spin connection}; $\mathcal{Q}$ - \emph{time torsion}; $T$ - \emph{space torsion}; $S$ - \emph{boost curvature}; $\Omega$ - \emph{space curvature}.

To achieve the UR limit of the Poincar\'e group, we rescale the group generators and fields as (See, for instance, reference \cite{Bergshoeff:2017btm})
\begin{eqnarray}
L_a&\longmapsto&\chi L_a\;,\nonumber\\
\Pi&\longmapsto&\chi\Pi\;,\nonumber\\
\theta^a&\longmapsto&\chi^{-1}\theta^a\;,\nonumber\\
q&\longmapsto&\chi^{-1} q\;.\label{res0}
\end{eqnarray}
These rescalings keep the first two expressions (fundamental fields decompositions in space and time) in \eqref{fdecomp1} unchanged. The UR limit is then achieved by $\chi\longrightarrow\infty$ at the Poincar\'e algebra \eqref{poincalg3}, at leading order. The result is the so called Carroll algebra,
\begin{eqnarray}
\left[L_{ab},L_{cd}\right]&=&\frac{1}{2}\left(\delta_{ad}L_{bc}-\delta_{ac}L_{bd}+\delta_{bc}L_{ad}-\delta_{bd}L_{ac}\right)\;,\nonumber\\
\left[L_{ab},\Pi_c\right]&=&\frac{1}{2}\left(\delta_{bc}\Pi_a-\delta_{ac}\Pi_b\right)\;,\nonumber\\
\left[L_{ab},L_c\right]&=&\frac{1}{2}\left(\delta_{bc}L_a-\delta_{ac}L_b\right)\;,\nonumber\\
\left[\Pi_a,L_b\right]&=&\frac{1}{2}\delta_{ab}\Pi\;,\label{carr1}
\end{eqnarray}
and zero otherwise. The contraction of the Poincaré algebra \eqref{poincalg3} to the Carroll one \eqref{carr1} is an example of the general procedure known as In\"on\"u-Wigner contraction \cite{Inonu:1953sp}. In the present case,  $ISO(1,3)\longrightarrow C(1,3)=SO(3)\times C(3)\times \mathbb{R}_s^3\times \mathbb{R}_t$ where $C(3)$ represents the Carrollian boosts \cite{Bergshoeff:2017btm}. The algebra \eqref{carr1} is clearly not semi-simple. Thence, it implies that Carrollian metrics are degenerate \cite{Ciambelli:2018ojf,Ciambelli:2019lap}. Moreover, relations \eqref{inv0} and \eqref{metrics0} imply on
\begin{eqnarray}
e^a_\mu e_b^\mu&=&\delta^a_b\;,\nonumber\\
q^\mu q_\mu&=&1\;,\nonumber\\
q^\mu e^a_\mu&=&q_\mu e_a^\mu\;=\;0\;,\nonumber\\
e^a_\mu e_a^\nu&=&\delta_\mu^\nu-q_\mu q^\nu\;.\label{inv1}
\end{eqnarray}

Besides the fact that $e^AP_A$ and $\omega^{AB}L_{AB}$ are kept unchanged under the rescalings \eqref{res0}, this is not true for the 2-form fields in \eqref{fdecomp1}. In fact, the 2-form fields will reduce to
\begin{eqnarray}
   T^a&=&De^a\;,\nonumber\\
\mathcal{Q}&=&dq-\frac{1}{2}\theta_a e^a\;,\nonumber\\
\Omega^a_{\phantom{a}b}&=&R^a_{\phantom{a}b}\;,\nonumber\\
S^a&=&D\theta^a\;.\label{fdecomp3}
\end{eqnarray}

For the gauge transformations  \eqref{gt0} one also needs to consider $\alpha^{AB}L_{AB}=\alpha_{ab}\Sigma_{ab}+\alpha^aG_a$ with $\alpha^a\longrightarrow\chi^{-1}\alpha^a$. Thus, Carroll gauge transformations attain the form, 
\begin{eqnarray}
\delta\omega^a_{\phantom{a}b}&=&D\alpha^a_{\phantom{a}b}\;,\nonumber\\
\delta\theta^a&=&D\alpha^a-\alpha^a_{\phantom{a}b}\theta^b\;,\nonumber\\
\delta e^a&=&-\alpha^a_{\phantom{a}b}e^b\;,\nonumber\\
\delta q&=&\frac{1}{2}e^b \alpha_b\;,\label{gt1}
\end{eqnarray}
Finally, one can easily check that UR limit of the hierarchy relations \eqref{hier0} is given by
\begin{eqnarray}
d\mathcal{Q}&=&\frac{1}{2}\left(\theta_aT^a-e_aS^a\right)\;,\nonumber\\
DR^a_{\phantom{a}b}&=&0\;,\nonumber\\
DT^a&=&R^a_{\phantom{a}b}e^b\;,\nonumber\\
DS^a&=&R^a_{\phantom{a}b}\theta^b\;.\label{hier1}
\end{eqnarray}

In the next sections we investigate some formal consequences of the UR limit of Lovelock-Cartan gravity, starting with the particular case of the EH action.

\section{Carroll gravity}\label{EH}

We start reviewing the usual UR limit of the EH action. The resulting gravity theory is typically known as \emph{Carroll gravity} \cite{Bergshoeff:2017btm,Ciambelli:2018ojf,Ciambelli:2019lap}.

\subsection{Action and field equations}

By imposing $\Lambda=z_1=z_2=z_3=z_4=0$ in the MZ action \eqref{mz1}, the EH action is obtained,
\begin{equation}
    S_{EH}=\kappa\int\epsilon_{ABCD}R^{AB}e^Ce^D+S_m\;.\label{eh0}
\end{equation}
The EH action \eqref{eh0}, in terms of the decompositions \eqref{fdecomp1} and \eqref{fdecomp2}, reads \cite{Bergshoeff:2017btm}
\begin{equation}
   S_{EH}=\kappa\int\epsilon_{abc}\left(2qR^{ab}e^c+\frac{1}{2}q\theta^a\theta^be^c-S^ae^be^c\right)+S_m\;,\label{eh1}
\end{equation}
with $\epsilon_{abc}=\epsilon_{\underline{0}abc}$. Rescaling the fields according to \eqref{res0}, together with $\kappa\longrightarrow\chi\kappa$, one gets
\begin{equation}
   S_{EH}=\kappa\int\epsilon_{abc}\left(2qR^{ab}e^c+\frac{1}{2}\chi^{-2}q\theta^a\theta^be^c-S^ae^be^c\right)+S_m\;.\label{ehaction2}
\end{equation}
and the UR limit is attained from $\chi\longrightarrow\infty$. The result is the Carroll gravity action\footnote{Obviously, the corresponding limit of the matter action $S_m$ must be consistent as well.} \cite{Bergshoeff:2017btm,Bergshoeff:2019ctr},
\begin{equation}
  S_C=\kappa\int\epsilon_{abc}\left(2qR^{ab}e^c-S^ae^be^c\right)+S_m\;.\label{carrollaction1}
\end{equation}
The field equations can be easily derived by varying the action \eqref{carrollaction1} with respect to $q$, $e$, $\omega$, and $\theta$ (in this order):
\begin{eqnarray}
\epsilon_{abc}R^{ab}e^c&=&-\frac{1}{2\kappa}\tau\;,\nonumber\\
qR^{ab}-\frac{1}{2}\left(S^ae^b-S^be^a\right)&=&\frac{1}{4\kappa}\epsilon^{abc}\tau_c\;,\nonumber\\
\mathcal{Q}e^a-qT^a&=&-\frac{1}{4\kappa}\epsilon^{abc}\sigma_{bc}\;,\nonumber\\
T^ae^b-T^be^a&=&\frac{1}{2\kappa}\epsilon^{abc}\sigma_c\;,\label{ehfeq1}
\end{eqnarray}
where we have defined
\begin{eqnarray}
\tau&=&\frac{\delta S_m}{\delta q}\;,\nonumber\\
\tau_a&=&\frac{\delta S_m}{\delta e^a}\;,\nonumber\\
\sigma_{ab}&=&\frac{\delta S_m}{\delta\omega^{ab}}\;,\nonumber\\
\sigma_a&=&\frac{\delta S_m}{\delta\theta^a}\;.\label{ehsour1}
\end{eqnarray}
The sources $\tau$ and $\tau_a$ are related to the relativistic energy-momentum tensor while $\sigma_{ab}$ and $\sigma_a$ with spin density. All sources defined in \eqref{ehsour1} are 3-forms fields.

One interesting feature of Carroll gravity is that it carries a Weyl symmetry which is not present in the EH gravity. In fact, considering the pure Carroll gravity action $S_{pC}=S_C-S_m$, one can easily check that it is invariant under a global scale transformation of the form
\begin{eqnarray}
e^a&\longmapsto&\mathrm{exp}(\zeta)e^a\;,\nonumber\\
q&\longmapsto&\mathrm{exp}(-\zeta)q\;,\nonumber\\
\theta^a&\longmapsto&\mathrm{exp}(-2\zeta)\theta^a\;.\label{weyl1}
\end{eqnarray}
with $\zeta$ being a global parameter. In functional form, the symmetry \eqref{weyl1} reads\begin{equation}
    \int\left(e^a\frac{\delta S_C}{\delta e^a}-q\frac{\delta S_C}{\delta q}-2\theta^a\frac{\delta S_C}{\delta\theta^a}\right)=0\;.\label{weyl2}
\end{equation}
This symmetry equips the fields $e$, $q$, $\theta$, and $\omega$ with a Weyl charge of $+1$, $-1$, $-2$, and $0$, respectively.   The fact that spin-connection is not Weyl-charged means that space curvature is invariant under Weyl transformations\footnote{The reason why the spin-connection can not be charged is because $R^{ab}$ carry linear and quadratic terms in $\omega^{ab}$. Thus, if $\omega^{ab}$ is charged, $R^{ab}$ would not transform with a global scale factor. Thence, because $R^{ab}$ appears explicitly in the action \eqref{carrollaction1}, any nontrivial charge would ruin Weyl symmetry.}. A local and simpler version of this symmetry is also present in the non-relativistic limit of gravity, but with different charges for the fields, see for instance \cite{Guerrieri:2020vhp,Devecioglu:2018apj}. Another difference is that in the present case, the field equations are complete since the boost connection appear in the action \eqref{carrollaction1} and also from the fact that we have a complete set of field equations in \eqref{ehfeq1}.

In order to specify the matter content with minimal requirements, we consider the class of matter actions obeying the Weyl symmetry \eqref{weyl2}. This class of matter actions encompass the usual Carroll limit of traditional actions such as the Klein-Gordon, Dirac and Maxwell actions, as one can verify by direct inspection from the results in \cite{Bergshoeff:2017btm}. Henceforth, it is not difficult to infer that $S_m$ must assume the more explicit general form
\begin{equation}
    S_m=\int\left[M_aqe^a+\Sigma\epsilon_{abc}\theta^ae^be^c+\frac{1}{2}\left(\pi_a\theta_b-\pi_b\theta_a\right)e^a e^b+\rho_{ab}\omega^{ab}\right]\;.\label{sm0}
\end{equation}
The densities  $M_a$, $\Sigma$, $\pi_a$, and $\rho_{ab}$ are allowed to depend only on $\omega^{ab}$. Moreover, $\Sigma$ and $\pi_a$ are 1-forms, $M_a$ is a 2-form, and $\rho_{ab}$ is a 3-form. Therefore, the quantities defined in \eqref{ehsour1} are now given by
\begin{eqnarray}
\tau&=&M_ae^a\;,\nonumber\\
\tau_a&=&-qM_a-2\Sigma\epsilon_{abc}\theta^be^c+\left(\pi_a\theta_b-\pi_b\theta_a\right)e^b\;,\nonumber\\
\sigma_{ab}&=&\rho_{ab}+\frac{\delta M_c}{\delta\omega^{ab}}qe^c+\frac{\delta\Sigma}{\delta\omega^{ab}}\epsilon_{cde}\theta^ce^de^e+\frac{\delta \pi_c}{\delta\omega^{ab}}\theta_d e^c e^d\;,\nonumber\\
\sigma_a&=&-\Sigma\epsilon_{abc}e^be^c-\pi_be^be_a\;.\label{ehsour2}
\end{eqnarray}

It is possible to extract some formal solutions from equations \eqref{ehfeq1}, provided \eqref{ehsour2}. For example, in vacuum, the trivial solution $R=S=T=\mathcal{Q}=0$ is accepted. Considering matter, in the form \eqref{ehsour2}, the first two equations in \eqref{ehfeq1} can be solved for $R^{ab}$ and $S^a$ to give 
\begin{eqnarray}
R^{ab}&=&-\frac{1}{4\kappa}\epsilon^{abc}M_c\;,\nonumber\\
S^a&=&\frac{1}{\kappa}\left(\Sigma\delta^a_c+\frac{1}{2}\epsilon^{ab}_{\phantom{ab}c}\pi_b\right)\theta^c\;.\label{RS1}
\end{eqnarray}
The second equation in \eqref{RS1} can be seen as a eigenvalue equation for the covariant derivative with $\theta^a$ being the eigenvectors and the 1-form $\frac{1}{\kappa}\left(\Sigma\delta^a_c+\frac{1}{2}\epsilon^{ab}_{\phantom{ab}c}\pi_b\right)$ the eigenvalues. Space torsion $T^a$ can be obtained from the fourth equation in \eqref{ehfeq1},
\begin{equation}
    T^a=-\frac{1}{2\kappa}\left(\Sigma\delta^a_c+\epsilon^{ab}_{\phantom{ab}c}\pi_b\right)e^c\;.\label{T1}
\end{equation}
Just like boost curvature, this equation can also be seen as an eigenvalue equation for the covariant derivative with $e^a$ as eigenvectors but with eigenvalues given by the 1-form $-\frac{1}{2\kappa}\left(\Sigma\delta^a_c+\epsilon^{ab}_{\phantom{ab}c}\pi_b\right)$. For the third equation in \eqref{ehfeq1} it is convenient to set\footnote{Without such imposition, $\mathcal{Q}$ cannot be easily isolated.} $\rho_{ab}=0$. Thence, one can isolate $\mathcal{Q}$,
\begin{equation}
    \mathcal{Q}=-\frac{1}{12\kappa}\left[q\left(6\Sigma-\epsilon^{abc}\frac{\delta M_c}{\delta\omega^{ab}}\right)+2\frac{\delta\Sigma}{\delta\omega^{ab}}\theta^ae^b+\frac{1}{2}\epsilon^{abc}\left(\frac{\delta \pi_c}{\delta\omega^{ab}}\theta_d-\frac{\delta \pi_d}{\delta\omega^{ab}}\theta_c\right)e^d\right]\;.\label{Q1}
\end{equation}
To obtain \eqref{Q1} the strategy is to isolate a space vierbein common to all terms. The symmetric part of the remaining terms imply on \eqref{Q1}, because it contains $\mathcal{Q}$. The antisymmetric part does not contain $\mathcal{Q}$ and gives the following relations
\begin{eqnarray}
\pi_a&=&-\frac{1}{4}\frac{\delta M^b}{\delta\omega^{ab}}\;,\nonumber\\
\frac{\delta\pi^b}{\delta\omega^{ab}}&=&0\;,\nonumber\\
\frac{\delta\Sigma}{\delta\omega^{ab}}&=&\frac{1}{4}\epsilon_{acd}\frac{\delta\pi^c}{\omega^{db}}\;.\label{mconst1}
\end{eqnarray}
These relations suggest (but do not imply) that $\Sigma$ and $\pi_a$ should not depend on the spin-connection. If so, time torsion \eqref{Q1} simplifies to
\begin{equation}
    \mathcal{Q}=-\frac{1}{12\kappa}q\left(6\Sigma-\epsilon^{abc}\frac{\delta M_c}{\delta\omega^{ab}}\right)\;.\label{Q1a}
\end{equation}

\subsection{Carroll-Riemann and Carroll- Weitzenb\"ock manifolds}

Two special geometries can be studied at formal level, namely, the Carroll-Riemann and the Carroll-Weitzenb\"ock geometries as solutions of the field equations \eqref{ehfeq1} for generic sources in the form \eqref{ehsour2}. The first one is defined by non-trivial curvatures and vanishing torsions. The second case is defined by vanishing curvatures and non-trivial torsions.

\subsubsection{Carroll-Riemann manifolds}

It is easy to infer, from \eqref{ehfeq1}, the conditions on the matter distributions in order to obtain a Carroll-Riemann geometry by setting $T^a=\mathcal{Q}=0$. The matter densities must satisfy then
\begin{eqnarray}
\Sigma\delta^a_c+\epsilon^{ab}_{\phantom{ab}c}\pi_b&=&0\;,\nonumber\\
\frac{\delta\Sigma}{\delta\omega^{ab}}+\frac{1}{4}\epsilon^{cde}\left(\frac{\delta \pi_e}{\delta\omega^{cd}}\delta_{ab}-\frac{\delta \pi_b}{\delta\omega^{cd}}\delta_{ae}\right)&=&0\;,\nonumber\\
6\Sigma-\epsilon^{abc}\frac{\delta M_c}{\delta\omega^{ab}}&=&0\;,\nonumber\\
\rho_{ab}&=&0\;.\label{CR1}
\end{eqnarray}
Taking the trace of the first condition of \eqref{CR1} imply on the vanishing of $\Sigma$ and $\pi_a$. Thence, conditions \eqref{CR1} reduce to
\begin{eqnarray}
\Sigma&=&0\;,\nonumber\\
\pi_a&=&0\;,\nonumber\\
\frac{\delta M_c}{\delta\omega^{ab}}&=&0\;,\nonumber\\
\rho_{ab}&=&0\;.\label{CR2}
\end{eqnarray}
We point out that such conditions are generic in such a way that no assumption on the fundamental gravitational fields is made. The corresponding curvatures read
\begin{eqnarray}
R^{ab}&=&-\frac{1}{4\kappa}\epsilon^{abc}M_c\;,\nonumber\\
S^a&=&0\;.\label{RS2}
\end{eqnarray}
Thus, the only non-trivial object is space curvature $R^{ab}$. Therefore, if $S_m$ depend only on $M_a$ and $M_a$ does not depend on any gravitational field, the resulting spacetime is a Carroll-Riemann manifold. Solutions \eqref{RS2} must fit to the Bianchi identities. In fact, from $DS^a=R^a_{\phantom{a}b}\theta^b$, one attains the possible solution $\theta^a=0$. From $DT^a=R^a_{\phantom{a}b}e^b$ and from the fact that space vierbein and space curvature are not vanishing quantities, we gain a constraint,
\begin{equation}
    \epsilon^{abc}e_bM_c=0\;.\label{constr00}
\end{equation}
Moreover, $DR^{ab}=0$ implies
\begin{equation}
    DM_a=0\;.\label{constr01}
\end{equation}

The fact that $\theta^a=0$ (see \eqref{fdecomp3}) implies on a solution for the time vierbein according to $dq=0\Rightarrow q=d\mathbf{t}$, with $\mathbf{t}$ being an arbitrary scalar function. In Galilean gravity, such function can be identified with absolute Newtonian time because $\oint d\mathbf{t}=0$. Hence, any clock would measure the same time interval, independently of the path observers take. Moreover, $q$ is a gauge independent quantity, ensuring the observational character of $\mathbf{t}$. In Carroll-Riemann geometry however, $q$ is not gauge invariant (see \eqref{gt1}). This property spoils the tempting interpretation of $\mathbf{t}$ being a kind of UR absolute time coordinate. At least, not before any gauge fixing. In fact, one can achieve the same conclusion (the absence of a gauge invariant absolute time coordinate) by setting $\mathcal{Q}=\theta^a=0$ to solve the field equations   \eqref{ehfeq1} in a more general approach.

\subsubsection{Carroll-Weitzenb\"ock manifolds}

For the Carroll-Weitzenb\"ock solutions, vanishing curvatures ($R^{ab}=S^a=0$) imply on the generic conditions (See \eqref{RS1}),
\begin{equation}
M_c=\Sigma=\pi_b=0\;.\label{M1}
\end{equation}
The corresponding torsions obtained from equations \eqref{ehfeq1} read
\begin{eqnarray}
    T^a&=&0\;,\nonumber\\
    \mathcal{Q}e^a&=&-\frac{1}{4\kappa}\epsilon^{abc}\rho_{bc}\;.\label{TQ1}
\end{eqnarray}
In this case, the only non-trivial object is time torsion. Moreover, the matter action $S_m$ must depend only on $\rho_{ab}$ which depends only on the spin connection. Due to the non-triviality of $\mathcal{Q}$, a UR absolute time definition is also out of question in Carroll-Weitzeinb\"ock manifolds.

Similarly to the Carroll-Riemann case, due to vanishing boost curvature, we can set $\theta^a=0$. Hence, all Bianchi identities are satisfied if the spin-density $\rho_{ab}$ obeys the constraint 
\begin{equation}
D\rho_{ab}=0\;.\label{constr02}
\end{equation}
Moreover, we can choose a Weitzenb\"ock-type connection $\omega^{ab}=0$. Thence, we end up with the set of equations  
\begin{eqnarray}
d\rho_{ab}&=&0\;,\nonumber\\
de^a&=&0\;,\nonumber\\
\mathcal{Q}&=&dq\;,\label{W1a}
\end{eqnarray}
to be solved together with the second of \eqref{TQ1}. The first equation in \eqref{W1a} says that the spin density can be written as an exact quantity, $\rho_{ab}=dX_{ab}$, with $X_{ab}$ being a 2-form. The second equation in \eqref{W1a} states that the space vierbein is also an exact form,
\begin{equation}
    e^a=dn^a\;,\label{e1}
\end{equation}
with $n^a$ being a 0-form. Thus, the second equation \eqref{TQ1} becomes
\begin{equation}
    d\left(dqn^a\right)=-\frac{1}{4\kappa}\epsilon^{abc}dX_{bc}\;\Rightarrow\;dqn^a=-\frac{1}{4\kappa}\epsilon^{abc}X_{bc}\;,
\end{equation} 
Defining $n^an_a=n^2$, we get
\begin{equation}
    q=-\frac{1}{4\kappa}\int\epsilon^{abc}\frac{n_aX_{bc}}{n^2}\;.\label{q1}
\end{equation}
Therefore, depending on the form of $X_{ab}$ and $n^a$, the final Weitzenb\"ock solution is given by the vierbeins \eqref{e1} and \eqref{q1} and vanishing connections.

\subsection{A non-trivial example}

A particularly interesting example appear if we set $\rho_{ab}=\Sigma=\pi_a=0$ in the field equations \eqref{ehfeq1} for the sources in the form \eqref{ehsour2}. Thence, $S^a=T^a=0$. Space curvature and time torsion remain non-trivial and read
\begin{eqnarray}
R^{ab}&=&-\frac{1}{4\kappa}\epsilon^{abc}M_c\;,\nonumber\\
\mathcal{Q}&=&\frac{1}{12\kappa}q\epsilon^{abc}\frac{\delta M_c}{\delta\omega^{ab}}\;.\label{RQ1}
\end{eqnarray}
Again, the fact that boost curvature vanishes allows to set $\theta^a=0$. Therefore,
\begin{equation}
\mathcal{Q}=dq\;,\label{Q00}
\end{equation}
and thus,
\begin{equation}
 dq=qf\;,\label{dq1}   
\end{equation}
with $f$ being the 1-form
\begin{equation}
f=\frac{1}{12\kappa}\epsilon^{abc}\frac{\delta M_c}{\delta\omega^{ab}}\;.\label{f1}
\end{equation}
For consistency, Bianchi identities \eqref{hier1} must be satisfied. Consequently, we need to impose \eqref{constr00}, \eqref{constr01}, and
\begin{equation}
    d(qf)=0\;.\label{constr03}
\end{equation}

Equation \eqref{dq1} can be easily solved for $q$ if we consider $f=\mathrm{constant}$, resulting in
\begin{equation}
    q=h\exp{(-f_\mu x^\mu)}+jf\;,\label{dq2}
\end{equation}
with $h$ being a constant 1-form and $j$ a constant 0-form. This is the same type of the solution found in \cite{Guerrieri:2020vhp} in Galilei gravity for the time torsion. Thus, following \cite{Guerrieri:2020vhp}, one can choose to work in ADM formalism in the temporal gauge, so $q=Ndt$ with $N$ being the lapse function. Moreover, without loss of generality, we can set $f=\mathbf{f}dt$, $h=\mathbf{h}dt$, and $j=\mathbf{f}^{-1}$. Thence,
\begin{equation}
    N(t)=\mathbf{h}\exp{(-\mathbf{f}t)}+1\;.\label{N1}
\end{equation}
Noting that, in this case, $qf=0\Rightarrow\mathcal{Q}=dq=0$ and a consistent foliation can be defined in such a way causality is ensured.

\begin{figure}[ht]
    \centering
    \includegraphics[width=0.8\textwidth]{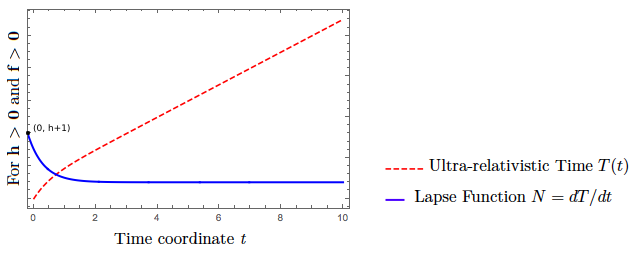}
    \caption{Illustration of the UR time $T(t)$ (red dashed line) and the lapse function $N(t)$ (blue solid line).}\label{fig1}
\end{figure}

The lapse function characterizes the rate between proper time $T$ and coordinate time $t$, namely $N=dT/dt$. Consequently, proper time $T(t)$ is given by
\begin{equation}
    T(t)=\frac{\mathbf{h}}{\mathbf{f}}\left[1-\exp{(-\mathbf{f}t)}\right]+t\;,\label{N2}
\end{equation}
where we have set $T(0)=0$. As the system evolves in time, we have $T(t)|_{t\rightarrow\infty}=t+\mathbf{h}/\mathbf{f}$ as $N(t)|_{t\rightarrow\infty}=1$. At this limit, proper time coincides with the coordinate time, up to a gap associated to time dilation. See Figure \ref{fig1}.

The conclusion here is that we have a spacetime with only nontrivial geometrical property given by space curvature and a foliation of space-like surfaces evolving in time. In \cite{Guerrieri:2020vhp}, a similar solution was found for $\mathcal{Q}$, but with nontrivial space torsion and vanishing curvatures.

\subsection{Spherically symmetric solution}\label{4.1}

Birkhoff theorem \cite{Wald:1984rg,Ryder:2009zz}, in GR, establishes that any spherically symmetric solution of Einstein equation will be static as well. We now proceed to check if such property survives the UR limit of the EH theory.

We have to solve the field equations \eqref{ehfeq1} within spherically symmetric imposition. For that, we consider the torsion free case, $T=\mathcal{Q}=0$, in vacuum, $\tau=\tau_a=\sigma_{ab}=\sigma_a=0$. Henceforth, the last two equations of \eqref{ehfeq1} are immediately satisfied. Consequently, $\omega^{ab}=\omega^{ab}(e)$ and $\theta^a=\theta^a(q,e)$. Therefore, the line element can be parameterized in the usual way \cite{Ryder:2009zz}
\begin{equation}
    ds^2 = - e^{2\alpha(t,r)}dt^2 + e^{2\beta(t,r)}dr^2 + r^2 d\Omega^2\;,\label{ds}
\end{equation}
where $d\Omega^2= d\theta^2 + sen^2\theta d\phi^2$ is the solid angle element. Considering the labels $u,v=\{2,3\}$, the corresponding curvatures read
\begin{eqnarray}
S^1 &=& 2\left(e^{-2\beta}\;[(\alpha'-\beta')\alpha' + \alpha''] + e^{-2\alpha}[(\dot{\alpha}-\dot{\beta})\dot{\beta} - \ddot{\beta}]\right) q e^1\;,\nonumber\\
S^v &=& \frac{2}{r}e^{-\beta}\left[e^{-\beta}\alpha' q + e^{-\alpha}\dot{\beta}e^1\right]e^v\;, \nonumber\\
R^{1v}&=& \frac{1}{r}e^{-\beta}d\beta \;e^v\;,\nonumber\\
R^{uv}&=& \frac{1}{r^2}(1 - e^{-2\beta})e^u e^v\;. \label{csscs}
\end{eqnarray}
The first two equations in \eqref{ehfeq1} decompose as
\begin{eqnarray}
\dot{\beta}&=&0\;, \nonumber\\
\frac{1}{2r^2}\left(1-e^{-2\beta}\right) + \frac{\beta'}{r}e^{-2\beta}&=&0\;, \nonumber\\
\frac{1}{2r^2}\left(1-e^{-2\beta}\right) - \frac{\alpha'}{r}e^{-2\beta}&=&0\;, \nonumber\\
\alpha'' + [\alpha' - \beta']\left(\alpha' + \frac{1}{r}\right)&=&0 \;. \label{eqcrrl}
\end{eqnarray}
The first equation in \eqref{eqcrrl} ensures that $\beta$ must not depend on time, $\beta=\beta(r)$. The remaining equations lead to the usual solution,
\begin{eqnarray}
\alpha&=&-\beta\;,\nonumber\\
e^{2\alpha} &=& 1 - \frac{2M}{r}\;. \label{eb}
\end{eqnarray}
Therefore,
\begin{equation}
    ds^2 = - \left(1 - \frac{2M}{r}\right)dt^2 + \left(1 - \frac{2M}{r}\right)^{-1}dr^2 + r^2 d\Omega^2\;, \label{schw}
\end{equation}
which coincides with the Schwarzschild solution of GR. The final form of the curvatures can be written as
\begin{eqnarray}
S^1 &=& \frac{4M}{r^3}\;q e^1\;,\nonumber\\
S^v &=& -\frac{2M}{r^3}\;qe^v\;, \nonumber\\
R^{1v}&=& \frac{M}{r^3}\;e^{1}e^v\;,\nonumber\\
R^{uv}&=& -\frac{2M}{r^3}\;e^u e^v\;. 
\end{eqnarray}
Therefore, we have verified that Birkhoff theorem holds for the UR limit of gravity described by Carroll geometry. Although expected, this result is not evident because Carroll action \eqref{carrollaction1} lacks from a piece of the original EH action \eqref{eh1}.

\section{Carroll-Cartan gravity}\label{GC}

In this section we study the UR limit of the MZ action \eqref{mz1}. We derive the corresponding field equations and provide some formal solutions.

\subsection{Action and field equations}

The MZ action action \eqref{mz1}, in terms of decompositions \eqref{fdecomp1} and \eqref{fdecomp3}, reads \cite{Guerrieri:2020vhp}
\begin{eqnarray}
    S_{MZ}&=&\kappa\int\left[\epsilon_{abc}\left(2qR^{ab}e^c+\frac{1}{2}q\theta^a\theta^be^c-S^ae^be^c\right)+2\Lambda\epsilon_{abc}qe^ae^be^c\right]+\nonumber\\
  &+&\int\left[z_1\left(R^{ab}e_ae_b-qe_aS^a-\frac{1}{4}e_a\theta^ae_b\theta^b\right)+z_2\left(Q^2+T^aT_a\right)\right]+\nonumber\\
    &-&\int\left[2z_3\epsilon_{abc}S^a\left(R^{bc}+\frac{1}{4}\theta^b\theta^c\right)-z_4\left(R^{ab}R_{ab}+\frac{1}{2}S^aS_a+\frac{1}{2}R^{ab}\theta_a\theta_b\right)\right]+S_m\;.\label{mz2}
\end{eqnarray}
Rescaling the fields according to \eqref{res0}, action \eqref{mz2} takes the form
\begin{eqnarray}
   S_{MZ}&=&\kappa\int\left[\epsilon_{abc}\left(2\chi^{-1} qR^{ab}e^c+\frac{1}{2}\chi^{-3}q\theta^a\theta^be^c-\chi^{-1}S^ae^be^c\right)+2\chi^{-1}\Lambda\epsilon_{abc}qe^ae^be^c\right]+\;,\nonumber\\
  &+&\int\left\{z_1\left(R^{ab}e_ae_b-\chi^{-2}qe_aS^a-\frac{1}{4}\chi^{-2}e_a\theta^ae_b\theta^b\right)+\right.\nonumber\\
  &+&\left.z_2\left[\chi^{-2}\left(dq\right)^2+De^aDe_a+\chi^{-2}q\theta^aDe_a+\chi^{-2}e_a\theta^adq+\frac{1}{4}\chi^{-2}e_a\theta^ae_b\theta^b\right]\right\}+\nonumber\\
&-&\int\left[2z_3\epsilon_{abc}\left(\chi^{-1}S^aR^{bc}+\frac{1}{4}\chi^{-3}S^a\theta^b\theta^c\right)-z_4\left(R^{ab}R_{ab}+\frac{1}{2}\chi^{-2}S^aS_a+\frac{1}{2}\chi^{-2}R^{ab}\theta_a\theta_b\right)\right]+\nonumber\\
    &+&S_m\;.\label{mz3}
\end{eqnarray}
For the coupling parameters, we consider the following rescalings
\begin{eqnarray}
\kappa&\longmapsto&\chi\;\kappa\;,\nonumber\\
\Lambda&\longmapsto&\chi^{-1}\Lambda\;,\nonumber\\
z_1&\longmapsto&z_1\;,\nonumber\\
z_2&\longmapsto&z_2\;,\nonumber\\
z_3&\longmapsto&\;z_3\;,\nonumber\\
z_4&\longmapsto&z_4\;.\label{res2}
\end{eqnarray}
Therefore, the UR limit of the MZ action \eqref{mz3} is achieved by taking $\chi\longrightarrow\infty$, at leading order. Thence,
\begin{equation}
  S_{CMZ}=\kappa\int\epsilon_{abc}\left(2qR^{ab}e^c-S^ae^be^c\right)+\int\left(z_1\;R^{ab}e_ae_b+z_2T^aT_a+z_4R^{ab}R_{ab}\right)+S_m\;,\label{cmz1}
\end{equation}
with $T^a$ given in \eqref{fdecomp3}. The resulting gravity theory will be called \emph{Carroll-Cartan gravity}. The action \eqref{cmz1} is easily interpreted: The first two terms are identical to the UR limit of the EH action, see \eqref{carrollaction1}; Terms in $z_1$ and $z_2$ are torsional and become topological if $z_2=-z_1$ (In that case, we have an UR version of the Nieh-Yan topological term); The term in $z_4$ is topological (The UR version of the Pontryagin term) and does not contribute to the field equations. Moreover, the UR limit of the matter action is assumed to be well behaved as well. 

By direct inspection, one easily finds that Weyl symmetry \eqref{weyl2} is lost. Nevertheless, it can be restored by considering the possibility of rescaling the parameters $z_1$ and $z_2$ by means of
\begin{eqnarray}
z_1&\longmapsto&\exp{(-2\zeta)}z_1\;,\nonumber\\
z_2&\longmapsto&\exp{(-2\zeta)}z_2\;.\label{zs1}
\end{eqnarray}
  Hence, by imposing Weyl symmetry again and from the fact that the matter action should not depend on the coupling $z_i$ parameters, we fall into the same class of matter actions of the form \eqref{sm0}. 

The field equations generated by the action \eqref{cmz1} are (see \eqref{ehsour1} and \eqref{ehsour2})
\begin{eqnarray}
\epsilon_{abc}R^{ab}e^c&=&-\frac{1}{2\kappa}\tau\;,\nonumber\\
qR^{ab}-\frac{1}{2}\left(S^ae^b-S^be^a\right) -\frac{1}{2\kappa}\left(z_1 + z_2 \right)\epsilon^{abc}R_{cd}e^d &=&\frac{1}{4\kappa}\epsilon^{abc}\tau_c\;,\nonumber\\
\mathcal{Q}e^a-qT^a+\frac{\left(z_1 + z_2\right)}{4\kappa}\epsilon^a_{\phantom{a}bc}D(e^be^c)&=&-\frac{1}{4\kappa}\epsilon^{abc}\sigma_{bc}\;,\nonumber\\
D(e^ae^b)&=&\frac{1}{2\kappa}\epsilon^{abc}\sigma_c\;.\label{feqmz1}
\end{eqnarray}
These equations differ from the EH case \eqref{ehfeq1} only by the terms in $(z_1+z_2)$. Moreover, the trivial vacuum solution $R=S=T=\mathcal{Q}=0$ is accepted as well. 

Similarly to the Carroll case, we can solve the field equations \eqref{feqmz1} outside a spherically symmetric object. Considering again vanishing torsions, $T=\mathcal{Q}=0$, in vacuum, $\tau=\tau_a=\sigma_{ab}=\sigma_a=0$, the last two equations in \eqref{feqmz1} are automatically satisfied. Considering again a spherically symmetric line element as in \eqref{ds} and curvature components as in \eqref{csscs}, we find that the first two equations in \eqref{feqmz1} can be rewritten exactly as \eqref{eqcrrl}. Hence, it naturally follows that Birkhoff's theorem also holds for Carroll-Cartan gravity.

\subsection{Solutions in the presence of matter}

We proceed in finding general formal solutions of equations \eqref{feqmz1}.

\subsubsection{Almost general solution}

  In the presence of matter, we consider a matter action in the form \eqref{sm0}. Such choice is consistent with the fact that we can always couple, to any gravity theory, the usual matter distributions that we can couple to the EH action.  Moreover, The extended Weyl symmetry \eqref{weyl1} and \eqref{zs1} is at our disposal to select matter actions of the form \eqref{sm0}. Thence, the matter densities appearing in equations \eqref{feqmz1} are given in \eqref{ehsour2}. 

First and fourth equations in \eqref{feqmz1} are exactly the same as the Carroll case \eqref{ehfeq1}. As a consequence, space curvature and space torsion do not change with respect to the Carroll case. Therefore,
\begin{eqnarray}
R^{ab}&=&-\frac{1}{4\kappa}\epsilon^{abc}M_c\;,\nonumber\\
T^a&=&-\frac{1}{2\kappa}\left(\Sigma\delta^a_c+\epsilon^{ab}_{\phantom{ab}c}\pi_b\right)e^c\;.\label{RT1}
\end{eqnarray}
The second equation in \eqref{feqmz1}, provided $R^{ab}$ in \eqref{RT1}, gives the boost curvature,
\begin{equation}
    S^a=\frac{1}{\kappa}\left(\Sigma\delta^a_c+\frac{1}{2}\epsilon^{ab}_{\phantom{ab}c}\pi_b\right)\theta^c-\frac{(z_1+z_2)}{4\kappa^2}M^a\;.\label{S1}
\end{equation}
The boost curvature \eqref{S1} differs from the Carroll case only by the term in $(z_1+z_2)$, as expected. Finally, time torsion can be obtained from the third equation in \eqref{feqmz1}, provided $T^a$ in \eqref{RT1}. Nevertheless, for simplicity, we set $\rho_{ab}=0$. Thence,
\begin{equation}
    \mathcal{Q}=-\frac{1}{12\kappa}\left[q\left(6\Sigma-\epsilon^{abc}\frac{\delta M_c}{\delta\omega^{ab}}\right)+2\frac{\delta\Sigma}{\delta\omega^{ab}}\theta^ae^b+\frac{1}{2}\epsilon^{abc}\left(\frac{\delta \pi_c}{\delta\omega^{ab}}\theta_d-\frac{\delta \pi_d}{\delta\omega^{ab}}\theta_c\right)e^d\right]-\frac{(z_1+z_2)}{12\kappa^2}\pi_ae^a\;,\label{Q2}
\end{equation}
which differs from the Carroll case \eqref{Q1} only by the term in $(z_1+z_2)$ -- also an expected result. Just like the Carroll case, the validity of \eqref{Q2} is subjected to some relations that must be satisfied, namely,
\begin{eqnarray}
\pi_a&=&-\frac{1}{4}\frac{\delta M^b}{\delta\omega^{ab}}\;,\nonumber\\
\pi_a&=&\frac{\kappa}{\left(z_1+z_2\right)}\left(2\frac{\delta\Sigma}{\delta\omega^{ab}}\theta^b+\epsilon^{abc}\frac{\delta\pi_c}{\delta\omega^{db}}\theta^d\right)\;,\nonumber\\
\Sigma&=&\frac{\kappa}{6\left(z_1+z_2\right)}\left(\frac{\delta\pi_a}{\delta\omega^{ab}}\theta^b+2\epsilon^{abc}\frac{\delta\Sigma}{\delta\omega^{ab}}\theta_c\right)\;.\label{rel2a}
\end{eqnarray}

\subsubsection{Carroll-Riemann and Carroll-Weitzenb\"ock manifolds}

The Carroll-Riemann geometry can be defined as the the particular case of null torsions $T=\mathcal{Q}=0$ and non-trivial curvatures. To see if such kind of geometries are accepted by the field equations \eqref{feqmz1} (for a generic $S_m$), one can set directly $T=\mathcal{Q}=0$ in the field equations. These conditions imply on the vanishing of the spin densities $\sigma_{ab}=\sigma_a=0$ in such a way that the third and fourth equations in \eqref{feqmz1} are identically satisfied. These conditions imply on the constraints \eqref{CR2} again. Hence, the non-trivial field strengths are given by
\begin{eqnarray}
R^{ab}&=&-\frac{1}{4\kappa}\epsilon^{abc}M_c\;,\nonumber\\
S^a&=&-\frac{(z_1+z_2)}{4\kappa^2}M^a\;.\label{RS3}
\end{eqnarray}
However, such solution is inconsistent with the Bianchi identities \eqref{hier1}, unless the conditions \eqref{constr00}, \eqref{constr01} and
\begin{equation}
\epsilon^{abc}\theta_bM_c=0\;,\label{constr04}
\end{equation}
are satisfied.

The Carroll-Weitzenb\"ock geometry would be obtained  by setting null curvatures, $R=S=0$, and considering non-trivial torsions. From \eqref{RT1} and \eqref{S1}, null curvatures lead to $M_a=\Sigma=\pi_a=0$. Hence, space torsion also vanishes. The only non-trivial field strength is then time torsion, which should also be determined from the second equation in \eqref{TQ1}. In fact, the analysis leading to the solutions \eqref{e1}, \eqref{q1} and vanishing connections holds in the present case as well.

\section{Conclusions}\label{FINAL}

In this work we have generalized Carroll theory of gravity by allowing the existence of torsional terms. With that purpose, we considered Mardones-Zanelli action in four dimensions and the corresponding UR limit. The resulting theory of gravity, called Carroll-Cartan gravity, generalizes the UR limit of EH gravity with additional torsional terms. For the sake of completeness, we first studied the UR limit of the EH. The main results obtained are listed below:
\begin{itemize}
    \item Carroll gravity (the UR limit of EH gravity) in the first order formalism was obtained. The action and the corresponding field equations are displayed in expressions \eqref{carrollaction1} and \eqref{ehfeq1}.
    
    \item   We found that Carroll gravity enjoys a global scale symmetry given by \eqref{weyl1}. Such useful symmetry was employed to select a special, yet quite general, form for the matter action, see \eqref{sm0}.
    
    \item A general formal solution in the presence of matter was found in \eqref{RS1}, \eqref{T1} and \eqref{Q1}. The validity of such solutions is subjected to certain conditions on the matter content, namely $\rho_{ab}=0$ and relations \eqref{mconst1}.
    
    \item By defining a Carroll-Riemann manifold as a solution of Carroll gravity with vanishing torsions and non-trivial curvatures, we were able to find the solution \eqref{RS1} subjected to the conditions \eqref{CR2}. Moreover, the Bianchi identities require that \eqref{constr00} and \eqref{constr01} hold. The fact that boost curvature vanishes allows the choice $\theta^a=0$. It implies that $\mathcal{Q}=dq=0\;\Rightarrow q=d\mathbf{t}\;$. In Newton-Cartan gravity \cite{Guerrieri:2020vhp}, this condition permits the definition of a Newtonian absolute time $\mathbf{t}$. Inhere, we argued that absolute time takes no place because $q$ is not a gauge invariant quantity. Henceforth, before any gauge fixing, $\mathbf{t}$ cannot be associated to an observational quantity. In fact, this argumentation always holds in any case where $\mathcal{Q}=\theta^a=0$.
    
    \item By defining a Carroll-Weitzenb\"ock manifold as a solution of Carroll gravity with vanishing curvatures and non-trivial torsions, we were able to find the solution \eqref{TQ1} subjected to the conditions \eqref{M1}. Bianchi identities enforce \eqref{constr02}. The choice of vanishing connections are consistently allowed. The set of remaining equations could be exactly solved for the vierbeins. The solutions are displayed in \eqref{e1} and \eqref{q1}, for an arbitrary 0-form $n^a$ and the specific spin density $\rho_{ab}$ given in \eqref{W1a}.
    
    \item A solution with non-trivial curvatures and torsions was developed. Boost curvature and space torsion were set to zero and the non-trivial space curvature and time torsion are given in \eqref{RQ1}. For a specific condition, we were able to obtain the lapse function \eqref{N1} and the proper time \eqref{N2} as a function of the coordinate time $t$ (See Figure \ref{fig1}). Thence, time dilation is an explicit effect found for this solution. Moreover, for the particular solution we choose, time torsion also vanishes. 
    
    \item Finally, we confirmed that Birkhoff's theorem remains valid in Carroll gravity. This is an expected result, but it is not trivial since the UR limit of the EH action lacks from a piece of the original EH action.
\end{itemize}

After this systematic study of the UR limit of the EH action, we proceed with the UR limit of the MZ action \eqref{mz1}. Our results are listed as follows:

\begin{itemize}
    \item Carroll-Cartan gravity \eqref{cmz1} was obtained from the UR of the MZ action \eqref{mz1}. The corresponding field equations were displayed in \eqref{feqmz1}.
    
     \item The global Weyl symmetry \eqref{weyl1} is not present anymore. However, it can be restored by extending the scale transformations to the parameters $z_1$ and $z_2$ by means of \eqref{zs1}.
     
     \item Since we expect that the matter content couple to gravity in the same way in both theories (EH and LC), we consider that the matter action remains in the form \eqref{sm0} again.
     
     \item Birkhoff's theorem is valid and was trivially verified.
     
     \item An almost general solution in the presence of matter was developed, see \eqref{RT1}, \eqref{S1}, and \eqref{Q2}. This solution generalizes the Carroll case obtained in \eqref{RS1}, \eqref{T1}, and \eqref{Q1}. In fact, the solution for space curvature and space torsion are the same. Time torsion and boost connection, however, carry contributions for the extra terms in the action \eqref{cmz1}.
     
      \item The existence of Carroll-Riemann manifolds (vanishing torsions and non-trivial curvatures) in the presence of matter was verified. The corresponding curvatures are proportional to $M^a$, see \eqref{RS3}. The Bianchi identities imply on the constraints \eqref{constr00}, \eqref{constr01}, and \eqref{constr04}.
      
      \item For Carroll-Weitzenb\"ock manifolds (vanishing curvatures and non-trivial torsions), we found that space torsion also vanishes and the only non-trivial field strength is the time torsion. In fact, the solution is exactly the same as the one found in Carroll gravity, \emph{i.e.}, vanishing connections and vierbeins given in \eqref{e1} and \eqref{q1}.
\end{itemize}

\section*{Acknowledgements}

This study was financed by The Coordena\c c\~ao de Aperfei\c coamento de Pessoal de N\'ivel Superior - Brasil (CAPES) - Finance Code 001.

\bibliographystyle{utphys2}
\bibliography{library}

\end{document}